\documentclass[twocolumn,prl,showpacs]{revtex4-1}
\usepackage{graphicx}
\usepackage{color}
\usepackage{amsmath}
\usepackage[normalem]{ulem}
\usepackage{verbatim}
\usepackage{amssymb}
\usepackage{esint}
\usepackage{hyperref}
\usepackage{siunitx}

\begin{document}

\title{Absolute absorption cross sections from photon recoil in a matter-wave interferometer}

\author{Sandra Eibenberger, Xiaxi Cheng, J.\,P.~Cotter, Markus Arndt}
\email[]{markus.arndt@univie.ac.at}
\affiliation{University of Vienna, Faculty of Physics, VCQ \& QuNaBioS, Boltzmanngasse 5, A-1090 Vienna, Austria}

\begin{abstract}
We measure the absolute absorption cross section of molecules using a matter-wave interferometer. A nanostructured density distribution is imprinted onto a dilute molecular beam through quantum interference. As the beam crosses the light field of a probe laser some molecules will absorb a single photon. These absorption events impart a momentum recoil which shifts the position of the molecule relative to the unperturbed beam. Averaging over the shifted and unshifted components within the beam leads to a reduction of the fringe visibility, enabling the absolute absorption cross section to be extracted with high accuracy. This technique is independent of the molecular density, it is minimally invasive and successfully eliminates all problems related to photon-cycling, state-mixing, photo-bleaching, photo-induced heating, fragmentation and ionization. It can therefore be extended to a wide variety of neutral molecules, clusters and nanoparticles.
\end{abstract}

\maketitle

Optical spectroscopy, essential in the early development of quantum theory, is now a ubiquitous tool in the natural sciences\,\cite{Demtroeder2003,Berkowitz2002}. In many cases it is desirable to access the electronic, vibrational or rotational properties of free molecules. Gas-phase spectroscopy has therefore been established in various systems, ranging from gas cells to free molecular beams or ion traps. Absorption measurements in vapor cells\,\cite{Coheur1996} yield relative spectral information -- i.e. the positions of absorption lines. However, this requires the analyte particles to be sufficiently volatile and to form a gas of sizable opacity. For many complex molecules gas phase data are lacking because the number densities are too weak. 
This holds in particular for non-fluorescent molecules. Gas phase spectra are therefore often derived from solvent analysis\,\cite{Hare1991}. In addition, measuring {\em absolute} values for absorption cross sections conventionally requires accurate knowledge of the vapor pressure. This is notoriously difficult to determine for complex molecules.

Combining absorption spectroscopy with light-induced fluorescence offers a higher signal-to-noise, and therefore sensitivity. This method has been refined to matrix-isolated samples down to the level of  single molecules\,\cite{Basche1992,Kastrup2004}. However, in free molecular beams the number of photons scattered by complex molecules is intrinsically limited. The fluorescent light is typically red-shifted with respect to the absorption wavelength, with the energy difference remaining in the molecules. Frequent repetition of the absorption cycle rapidly leads to heating, photo-bleaching or destruction of the particles. This problem can be alleviated by keeping the analyte particles in a dilute buffer gas, which serves as a heat bath. This has been successfully exploited with isolated trapped biomolecular ions\,\cite{Rizzo2009,Antoine2011} and with molecules isolated in cryogenic matrices\,\cite{Dvorak2012}.

In this Letter we present a method to measure absolute absorption cross sections which circumvents repeated photon-cycling and which can be applied to extremely dilute beams. We exploit photon recoil in a Kapitza-Dirac-Talbot-Lau (KDTL) matter-wave interferometer\,\cite{Gerlich2007} to measure the reduction in quantum interference contrast as a function of the position and intensity of a recoil laser. From this we extract the absorption cross section $\sigma_{\mathrm{abs}}(\lambda_{k})$ at the laser wavelength $\lambda_{k}$\,\cite{Nimmrichter2008}.

\begin{figure}[t!]
\centering
\includegraphics[width = 1 \columnwidth]{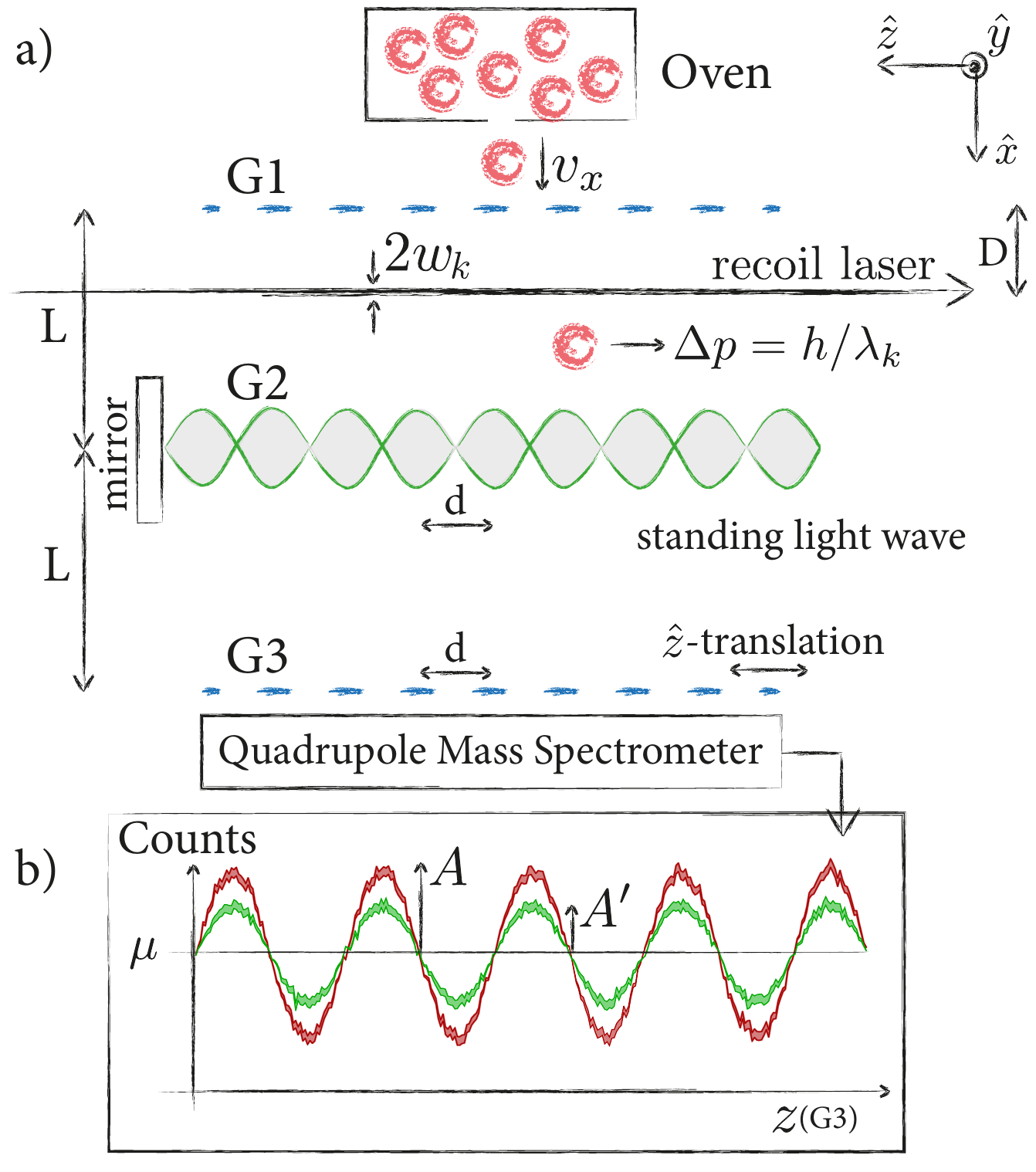}
\caption{Absolute absorption spectroscopy in the KDTL matter-wave interferometer. a) Grating $G1$ produces a spatially coherent source of molecules from the thermal beam emitted by an oven. After a distance $D$ the molecules encounter the recoil laser where the probability of absorbing $n$ photons, of wavelength $\lambda_{k}$, is described by a Poissonian distribution, $P_{n}$. At a distance $L-D$ further the molecules diffract at the standing light wave $G2$. The resulting interference pattern is read out by translating grating $G3$ laterally before a quadrupole mass spectrometer. b) The effect of some molecules absorbing a photon from the recoil laser results in a reduction of the observed visibility $\mathcal{V} = A/\mu \rightarrow \mathcal{V}^{\prime}=A^{\prime}/\mu$ when averaged over the molecular velocity distribution. Here $A$ (and $A^{\prime}$) are the unperturbed (perturbed) amplitude and $\mu$ is the mean of the near-sinusoidal quantum interference curve.
\label{fig:setup}}
\end{figure}
The delocalization of complex molecules has been studied in a variety of experimental arrangements, from far-field diffraction to near-field interferometry\,\cite{Hornberger2012}. The KDTL interferometer has enabled  observation of quantum interference with the most massive nanoparticles to date, exceeding $10^{4}$\,AMU\,\cite{Eibenberger2013}. The interference fringes in the KDTL interferometer are free flying periodic nanostructures formed by the molecular number density distribution. If this nanostructure is exposed to a uniform external force the internal particle properties determine the resulting shift of the interference pattern. This method has enabled measurements of electric polarizabilities\,\cite{Berninger2007}, susceptibilities and dipole moments\,\cite{Eibenberger2011}, thermally induced conformational dynamics\,\cite{Gring2010} as well as the distinction of structural isomers\,\cite{Tuexen2010}. Here we consider the case of photon recoils associated with the Poissonian statistics of a molecule either absorbing, or not absorbing, a photon. In particular where internal conversion dissipates the photon energy and suppresses any secondary emission. Instead of shifting the interference pattern this has the effect of reducing the observed interference contrast. Photon-induced decoherence, caused by the random emission of spontaneous or thermal photons has previously been studied in atom\,\cite{Chapman1995, Pfau1994} and molecule\,\cite{Hackermueller2004} interferometry, respectively. Photon absorption in Ramsey-Bord\'{e} interferometry\,\cite{Borde1989} has also previously been used to determine the fine structure constant\,\cite{Weiss1993} and molecular data. Here we make use of fringe averaging in KDTL interferometry to derive precise absorption spectra, as suggested in reference\,\cite{Nimmrichter2008}.

A schematic of the experimental setup is shown in Fig.~\ref{fig:setup}a). In this particular demonstration, molecules of the fullerene C$_{70}$ are sublimated in a thermal source at a temperature of $654\,$\si{\degree}C and fly through three gratings, $G1$, $G2$ and $G3$ before detection in a Quadrupole Mass Spectrometer (QMS). Gratings $G1$ and $G3$ are etched into SiN with a period $d = 266\,$nm and an opening fraction of $0.42$. Grating $G2$ is a standing light wave formed by a laser with wavelength $\lambda_{g} = 532\,$nm and waists $w_x\simeq 20\,\mathrm{\mu m}$ and $w_y\simeq 500\,\mathrm{\mu m}$. Adjacent gratings are spaced by a distance $L \simeq 10.5\,$cm\,\cite{Gerlich2007}. The collimation and velocity selection of the molecular beam is provided by a series of slits. This results in a beam height of about $200\,\mathrm{\mu}$m and a divergence less than $1\,$mrad. The velocity distribution is measured using a time-of-flight technique using a rotating chopper disc and time-resolved detection\,\cite{Comsa1981}. We find the velocity distribution to be well approximated by $P(v_{x}) = (\sqrt{2 \pi} \sigma_{v_{x}})^{-1} \exp{\left[-(v_{x}-v_{0})^{2} / 2 \sigma_{v_{x}}^{2}\right]}$ with a mean velocity $v_{0} = 210.3(7)\,$ms$^{-1}$ and standard deviation $\sigma_{v_{x}} = 38.4(5)\,$ms$^{-1}$. Here quoted errors are the $1\sigma$ uncertainty estimates. The molecular beam is initially incoherent. However, the openings of $G1$ represent a comb of slits which act as narrow and therefore spatially coherent sources of transmitted wavelets\,\cite{Hornberger2012}. The molecules interact with $G2$ through the optical dipole force. The spatially periodic variation of the laser intensity imprints a spatially varying phase onto the transmitted matter wave, which evolves into a modulation of the molecular density distribution. The QMS then counts the number of molecules transmitted by G3. The molecular density pattern arriving at $G3$ has a near-sinusoidal shape. This is revealed in the count rate when $G3$ is laterally translated as illustrated in Fig.~\ref{fig:setup}b). The visibility of this curve $\mathcal{V} =A/\mu$ is determined by the fringe mean $\mu$ and amplitude $A$. For the experiments described here count rates of $\mu \simeq 300\,$/s and a fringe visibility of $\mathcal{V}\simeq 0.15$ were typical. Decoherence, dephasing or phase averaging in the interferometer reduce the fringe visibility $\mathcal{V}$. Under controlled conditions this reduction can be used to extract detailed information about the interaction between the molecules and their environment.

Consider a particle which absorbs $n$ optical photons from a recoil laser whose k-vector is parallel to the grating vector. The laser is positioned at a distance $D$ from $G1$, where $D<L$. If a molecule absorbs a photon it receives a momentum recoil $\Delta p = h/ \lambda_{k}$. In the plane of $G3$ this recoil shifts the position of the absorbing molecule relative to the unperturbed interference pattern by a distance $s = \Delta p D / m v_{x} = \lambda_{dB} D / \lambda_{k}$. The probability of a molecule absorbing $n$ photons while crossing the recoil laser is described by the Poissonian distribution $P_{n}(n_{0})=n_{0}^{n}\exp{(-n_{0})}/n!$ where the average number of absorbed photons $n_0$ is obtained by integrating over the Gaussian recoil laser intensity profile along the $\hat{x}$-axis
\begin{equation}
n_{0} = \sqrt{\dfrac{2}{\pi}}~\dfrac{\sigma_{\mathrm{abs}}\,\lambda_{k}\,P_{k}}{h\,c\,w_{k,y}\,v_{x}}.
\label{eq:n0}
 \end{equation}
A relative dephasing within the total ensemble occurs because not all molecules acquire the same momentum recoil. For a given molecular velocity $v_{x}$ this results in a reduced interference contrast $\mathcal{V}^{\prime} = R \mathcal{V}$ where\,\cite{Nimmrichter2008} $R = \exp{\left( -n_{0} \left[ 1 - \cos{\left(2 \pi s/d\right)} \right] \right)}$. However, for a realistic molecular beam with a finite spread in velocities the observed reduction in contrast must be averaged over the velocity distribution. This results in the relation, $\mathcal{V}^{\prime} = \langle R \rangle_{v_{x}} \mathcal{V}$, where the reduction in visibility is now described by\,\cite{Nimmrichter2008}
\begin{eqnarray}
\langle R \rangle_{v_{x}} &=&  \left| \int_{0}^{\infty} dv_{x} P(v_{x}) \exp{\left( -n_{0}\left[ 1 - \exp\left( \dfrac{2 \pi i s}{d} \right) \right] \right)}\right|. \nonumber \\ ~ & ~
\label{eq:R2}
\end{eqnarray}

\begin{figure}[t!]
\centering
\includegraphics[width = \columnwidth]{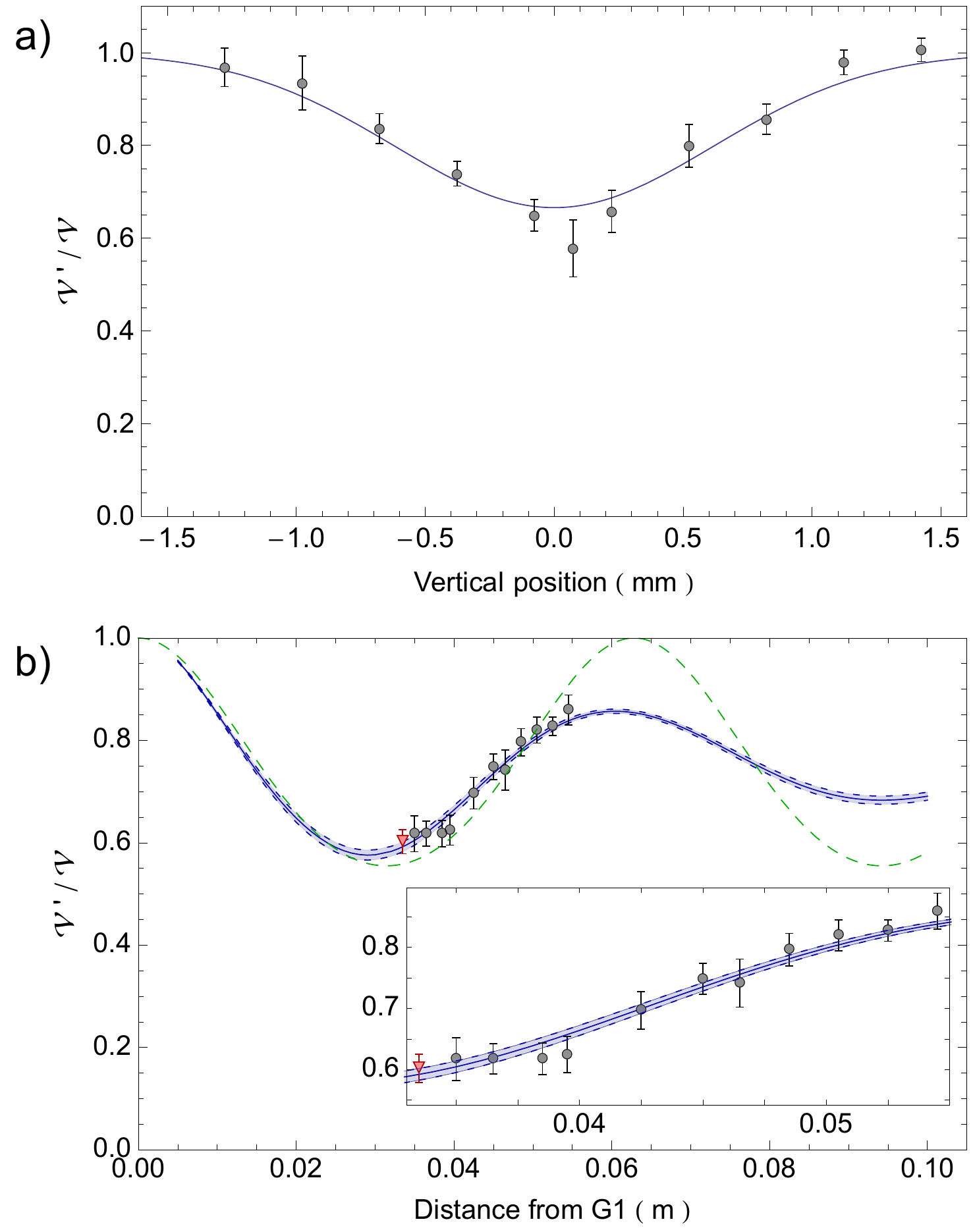}
\caption{a) Reduction in interference fringe visibility as a function of the vertical position of the recoil laser for a fixed $D = 3.5\,$cm. The data show the mean value from $5$ runs and their standard deviation. The solid line is a Gaussian fit with a fixed waist of $1.23\,$mm. b) Reduction in interference contrast as a function of the distance $D$ between recoil laser and $G1$. The data show the mean value from $10$ runs and their standard errors. The solid blue line describes the visibility reduction factor $\langle R \rangle_{v_{x}}$ for the absorption cross section $\sigma_{\mathrm{abs}}$ which minimizes the $\chi^{2}$. The shaded blue region shows $\langle R \rangle_{v_{x}}$ between the $1\sigma$ estimates of $\sigma_{\mathrm{abs}}$.
The dashed green line is the reduction factor $R$ for a monochromatic molecular beam, $P(v_{x}) = \delta(v_{x}-v_{0})$. The data point highlighted as a red triangle corresponds to the visibility reduction when the recoil laser is closest to $D_{\mathrm{min}}^{(1)} = d m v_{0} \lambda_{k}/2 h$. In a simplified version of our analysis this is used to estimate the absorption cross section.
\label{fig:reduction}}
\end{figure}

In our experiments the height of the molecular beam is small compared to the waist of the recoil laser which was measured using a scanning knife-edge profiler to be $w_{kx} =w_{ky} = 1.23(2)\,$mm. Fig.~\ref{fig:reduction}a) shows the variation of $\mathcal{V}^{\prime}/\mathcal{V}$ when the recoil laser is translated along the $\hat{y}$-axis. We find good agreement with the waist measured using the profilometer and the contrast reduction.

In order to measure $\sigma_{\mathrm{abs}}$ we record $\mathcal{V}^{\prime}/\mathcal{V}$ for a variety of different horizontal laser positions $D$. The current experimental arrangement allows for optical access between $3.5 < D < 5.5\,$cm. The recoil laser is fixed at a power of $P_{k} = 17.4(2)\,$W and a wavelength of $\lambda_{k} = 532.2\,$nm, with a drift on the order of $10\,$MHz over the course of our measurements.
Fig.~\ref{fig:reduction}b) shows how the observed reduction in fringe visibility varies with $D$. From a fit of equation (\ref{eq:R2}) to this data we determine the absorption cross section to be
\begin{equation}
\sigma_{\mathrm{abs}}(532\,\mathrm{nm}) = 1.97(6)\times 10^{-21}\,\mathrm{m}^{2},
\label{eq:absMeas}
\end{equation}
without any other free parameter. The quoted uncertainty in $\sigma_{\mathrm{abs}}$ is the $1\sigma$ estimate derived from the variation of the $\chi^{2}$ for different values of $\sigma_{\mathrm{abs}}$. The systematic uncertainty is dominated by knowledge of the recoil laser power and waist, which for these experiments enters at the level of $4\times10^{-23}\,$m$^{2}$.

A faster method to measure the absorption cross section can be employed with similar precision and only slightly reduced accuracy. For a fixed velocity $P(v_{x}) = \delta(v_{x} - v_{0})$ the first minimum in the reduction of contrast occurs at $D_{\mathrm{min}}^{(1)} = d m v_{0} \lambda_{k}/2 h$ and the absorption cross section is described by
\begin{equation}
\sigma_{\mathrm{abs}} = - \sqrt{\dfrac{\pi}{8}} \dfrac{h\,c\,w_{y}\,v_{x}}{P_{k}\,\lambda_{k}}\ln{(\mathcal{V}^{\prime}/\mathcal{V})}.
\label{eq:absVdelta}
\end{equation}
Close to $D = D_{min}^{(1)}$ the effect of finite velocity spread is small for all values of $v_{0}$. The data point highlighted as a red triangle in Fig.~\ref{fig:reduction}b) is the nearest to this minimum we could achieve in our current setup, with a position $D = 3.5$\,cm. If we ignore the effects of velocity spread and use equation (\ref{eq:absVdelta}) to determine the absorption cross section we find a value $\sigma_{\mathrm{abs}} = 1.7\times10^{-21}\,$m$^{2}$, which underestimates the value determined from our detailed analysis by approximately $14\%$. This type of measurement is subject to systematic shifts towards lower absorption cross sections. However, this effect is small compared to the change in absorption cross section close to molecular resonances which can span decades. A further simplification can be made for molecules with a very low mean velocity or for short recoil laser wavelengths. The spatial period of $\mathcal{V}^{\prime}/\mathcal{V}$ is $2 D_{\mathrm{min}}^{(1)} = d m v_{0} \lambda_{k}/h$ for a monochromatic beam, varying only slightly when realistic velocity distributions are considered. When $D$ is very large compared to $2 D_{\mathrm{min}}^{(1)}$ the reduction factor is described by $\langle R \rangle_{v_{x}} \simeq \exp \left( -n_{0} \right)$ and the sensitivity to $D$ is removed.

\begin{figure}[htb]
\centering
\includegraphics[width = \columnwidth]{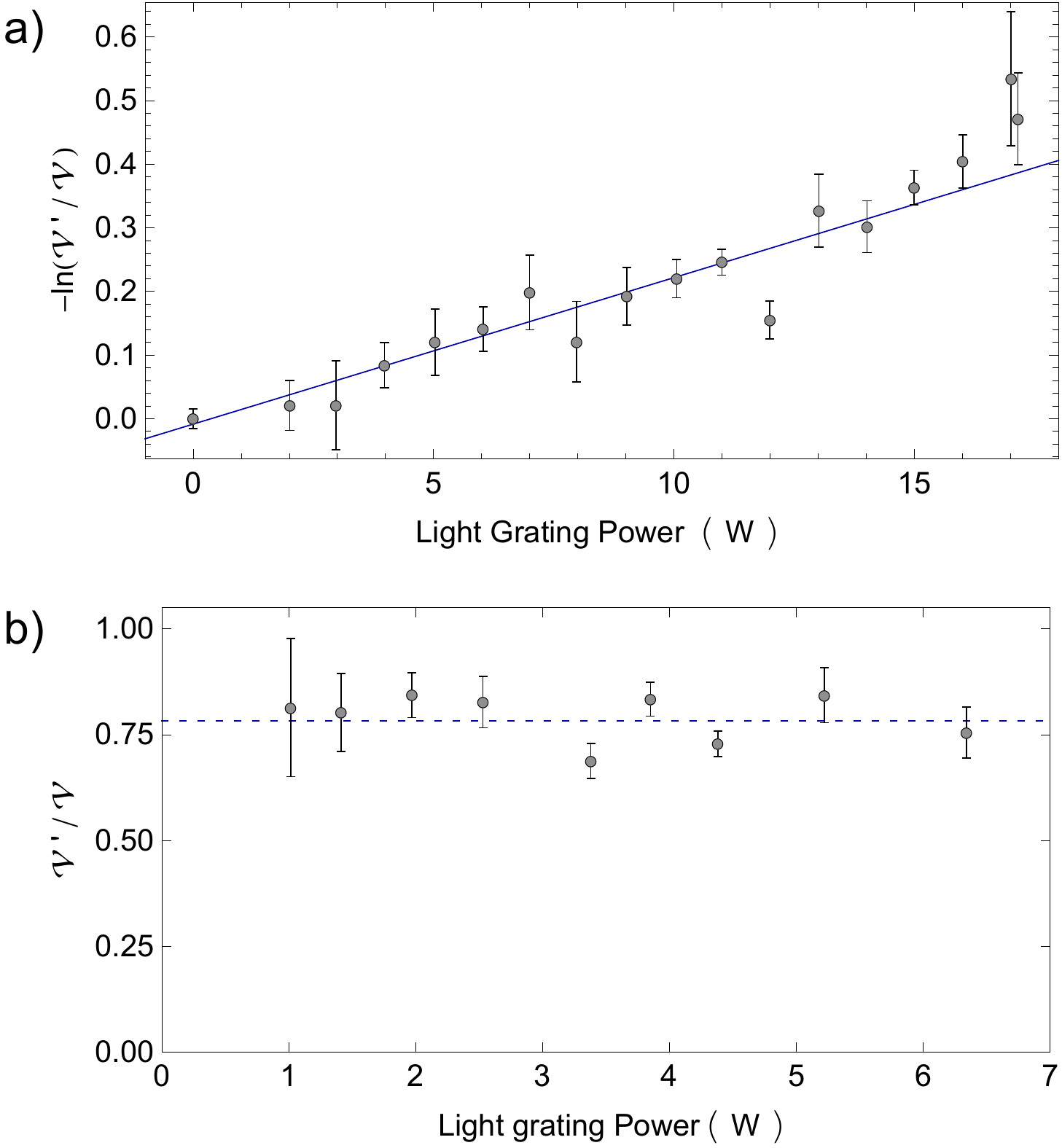}
\caption{a) Dependence of $-\ln(\mathcal{V}^{\prime}/\mathcal{V})$ on the recoil laser power for a distance $D=3.5\,$cm, as marked with the red data point in Fig.~\ref{fig:reduction}b). b) The visibility reduction $\mathcal{V}^{\prime}/\mathcal{V}$ as a function of the light grating power for a recoil laser power of $P_k = 17.4(2)\,$W and a distance $D=4.85\,$cm. We see a uniform value of $\mathcal{V}^{\prime}/\mathcal{V}=0.78(8)$ for light grating powers ranging from $1 - 6.5 \,$W. For the time scales relevant to our setup we find that the excited state polarizability plays no role for the reduction in fringe visibility.
\label{fig:pol}}
\end{figure}

Our model assumes that multi-photon processes are negligible. To ensure this is valid we recorded the reduction in contrast for recoil laser powers ranging from $0 - 17.4$\,W. Fig.~\ref{fig:pol}a) shows a clear linear relation between $-\ln(\mathcal{V}^{\prime}/\mathcal{V})$ and the recoil laser power. This behaviour is expected from the definition of $R$ and we therefore determine a maximum mean absorbed photon number of $0.14$.

The interaction potential at $G2$ depends on the optical polarizability of the molecules. Although they enter the interferometer in the electronic ground state S$_1$ a small fraction can be excited to the triplet state T$_1$ by the recoil laser. If some remain in T$_1$ at $G2$ this will be seen as a change in $\mathcal{V}^{\prime}/\mathcal{V}$ as the light grating power varies. For a distance of $(L - D) \simeq 5\,$cm and a velocity of $\simeq 210\,$ms$^{-1}$ the molecules take approximately $240\,\mu$s to reach $G2$ after the recoil laser. This is significantly longer than the expected triplet lifetime of C$_{70}$ in the gas phase which has been measured to be $\tau \simeq  41\,\mu$s\,\cite{Haufler1991}. Fig.~\ref{fig:pol}b) shows the results of experiments where we measured $\mathcal{V}^{\prime}/\mathcal{V}$ for light grating powers ranging from $1 - 6.5\,$W. The recoil laser here is positioned $4.85\,$cm from $G1$ to ensure a reasonable reduction in contrast while maintaining a good signal-to-noise ratio at low light grating powers. We see no significant change in $\mathcal{V}^{\prime}/\mathcal{V}$ and conclude that the excited triplet state plays no significant role in our absorption cross section measurements.

In summary, we have demonstrated a precise and accurate method to measure the absolute absorption cross section of molecules in dilute beams from quantitative observations of the fringe visibility as a function of the intensity and position of a probe laser. Our result for C$_{70}$ at 532\,nm is in good agreement with previous absorption experiments in a vapor cell $\sigma_{abs} \simeq 1.3 - 2.3\times10^{-21}\,$m$^{2}$\,\cite{Coheur1996}, an extrapolation from a thin film measurement $\sigma_{abs} \simeq 3.4\times10^{-21}\,$m$^{2}$\, \cite{Yagi2009} and from an indirect measurement using molecular interferometry $\sigma_{abs} = 2.5\times10^{-21}\,$m$^{2}$\,\cite{Hornberger2009}. We realize an absolute uncertainty of $3\%$ which can be used to anchor relative spectra to our fixed frequency value at 532\,nm with the same absolute accuracy. Further improvements are conceivable, especially for beams with a narrower velocity spread. At $T=654$\,\si{\degree}C several hundred rotational and vibrational states are populated in C$_{70}$. We have therefore not included the role of temperature in our analysis. It will be interesting to extend our studies in the future to low temperatures where some or all vibrational degrees are frozen out.

Our method can be applied to many molecular species and is of particular interest to extremely dilute beams whose vapor pressures cannot be known a priori. This applies especially to neutral biomolecular beams which could be studied as isolated particles or embedded in nanosolvent environments. One may also consider multi-photon experiments with higher time resolution for various interferometer configurations\,\cite{Hornberger2012,Haslinger2013}.

Absorption spectroscopy through photon recoil in matter-wave interferometry will be important in future research at the interface between quantum optics, physical chemistry and biomolecular physics\,\cite{Vries2007}.

We acknowledge valuable discussions with S. Nimmrichter and K. Hornberger. We are grateful for financial support from the FWF through projects Z149-N16 and DK CoQuS W1210-2, the European commission through project ERC AdvG PROBIOTIQUS(320694) and the Vienna ZIT communication project (957475). JPC is supported by a VCQ fellowship.

\bibliographystyle{apsrev4-1}
%


\end{document}